\documentstyle[12pt,epsf]{article}
\newcommand{\email}[1]{\texttt{email: #1}}
\newcommand{\bsigma}{\mbox{\boldmath $\sigma$}}

\begin{document}

\title{\bf Density-matrix spectra for integrable models}
 
\author{
Ingo Peschel\thanks{\email{peschel@physik.fu-berlin.de}}
\, and Matthias Kaulke\thanks{\email{kaulke@physik.fu-berlin.de}} \\
{\small Fachbereich Physik, Freie Universit\"at Berlin,} \\
{\small Arnimallee 14, D-14195 Berlin, Germany} \\
 \\
 \"Ors Legeza\thanks{\email{olegeza@power.szfki.kfki.hu}}\\
{\small Research Institute for Solid State Physics} \\
{\small P.O. Box 49, H-1525 Budapest, Hungary}
}
 
\maketitle
 
\begin{abstract}
The spectra which occur in numerical density-matrix renormalization
group (DMRG) calculations for quantum chains can be obtained
analytically for integrable models via corner transfer matrices. This
is shown in detail for the transverse Ising chain and the uniaxial
$X\!X\!Z$ Heisenberg model and explains in particular their exponential
character in these cases.

\medskip

\end{abstract}

\newpage

\section{Introduction}
The density-matrix renormalization group (DMRG) introduced by White in
1992 \cite{white} is a numerical procedure by which one-dimensional
quantum systems can be treated with spectacular accuracy. It is not
difficult to obtain the ground-state energy for spin chains with 100
sites up to nine decimal places \cite{legeza}. Correlation functions
can be calculated as well and the DMRG has therefore become an
important new tool in this area of physics.

The basis of the method is a proper selection of states in the Hilbert
space which are important for the target state one wants to
study. This is done by dividing the system into two parts with
corresponding reduced density matrices $\varrho_1$ and $\varrho_2$. The
relevant states in each part are then given by those eigenvectors of
the $\varrho_\alpha$ which have the largest eigenvalues. Thus the spectra
of the $\varrho_\alpha$ enter in an essential way and it is obvious that
the method will only work well if the eigenvalues drop rapidly enough
so that a small number of states is sufficient and practically
exhausts the sum rules ${\rm Tr}(\varrho_\alpha )=1$. Spectra of such a
form, where the eigenvalues decrease roughly exponentially, have
indeed been observed in various calculations
\cite{white,kaulke}. Somewhat surprisingly, however, no detailed
investigation or discussion of their origin has been given so far. In
this article we want to present such a study for two particular
systems, the Ising chain in a transverse field and the $X\!X\!Z$
Heisenberg chain.

Although these systems are integrable and their ground states are
known, a direct determination of the corresponding $\varrho_\alpha$ is
difficult. Therefore, we first relate the quantum chains to
two-dimensional classical systems, namely the Ising model and the
six-vertex model. As pointed out by Nishino et al. \cite{nishino}, the
density matrices $\varrho_\alpha$ then become partition functions of
strips with a cut and these can, in turn, be expressed as products of
corner transfer matrices (CTM's) \`a la Baxter \cite{baxter}. This
feature {\em per se} is quite general. For integrable models, however,
the spectra of such CTM's are known in the thermodynamic limit and
have, in fact, exponential form, i.e. $\omega_n\sim\exp(-\alpha\, n)$
with integer $n$. Provided the correlation length $\xi$ of the system
is much smaller than the strip width or chain length $L$, the same
should then hold for the $\varrho_\alpha$. This is, indeed, the
case: The spectra obtained from DMRG calculations are found to agree
very well with the CTM results, both in their exponential form and in
the predicted degeneracies. Deviations occur only at the lower end
where finite-size and geometry effects are visible. In this way one
obtains a simple and consistent picture of the density-matrix spectrum
and its origin.

In the following we first describe the relation between density matrix
and CTM and recall the results available for the latter. Then we
present the numerical calculations for the two chains and the
comparison with the analytical predictions. The conclusion contains
some further discussion, also with respect to non-integrable and to
critical systems.

\section{Density matrices and corner transfer matrices}
Consider a spin one-half quantum chain with $L$ sites and Hamiltonian
$H$. The density matrix constructed from the ground state
$|\Phi\rangle$ is
\begin{equation}
\varrho=|\Phi\rangle \langle\Phi|
\end{equation}
or in a matrix notation, with $\bsigma=\{
\sigma_1,\sigma_2,\ldots ,\sigma_L\}$ denoting a spin configuration of
the chain and assuming $\Phi$ to be real,
\begin{equation}
\varrho(\bsigma,\bsigma')=\Phi(\bsigma)\Phi(\bsigma'). 
\end{equation}

By taking partial traces one obtains the reduced density matrices
which are of interest in the following. Writing
$\bsigma_1=\{\sigma_1,\sigma_2,\ldots ,\sigma_M\}$ and
$\bsigma_2=\{\sigma_{M+1},\sigma_{M+2},\ldots ,\sigma_L\}$ one has
\begin{equation}
\varrho_1(\bsigma_1,\bsigma_1')=\sum_{\bsigma_2}\Phi(\bsigma_1,\bsigma_2)
\Phi(\bsigma_1',\bsigma_2) 
\end{equation}
and similarly for $\varrho_2$.

Now imagine that there is a relation of the quantum chain to a two-dimensional
lattice of classical spins. There are various possibilities for
this. For example, $H$ could be a derivative of the row-to-row
transfer matrix $T$ of the lattice, or $T$ could arise from a Trotter
decomposition of $\exp(-\beta H)$. Here we want to assume that the
commutation relation
\begin{equation}
\left[ H,T \right] =0
\end{equation}
holds. Then the ground state $|\Phi\rangle$ of $H$ is also an eigenstate of
$T$, and if it also gives the maximal eigenvalue (which is the case in
our examples), it can be obtained from an arbitrary starting vector by
applying $T$ a large number of times. Therefore $\Phi(\bsigma)$ can be
viewed as the partition function (properly normalized) of a
half-infinite vertical strip with the spin configuration $\bsigma$ at
the near end and an arbitrary one at the other, far away
end. Similarly, $\varrho(\bsigma,\bsigma')$ can be regarded as the
partition function for two such strips, one extending to $-\infty$ and
the other one to $+\infty$, with end configurations $\bsigma$ and
$\bsigma'$, respectively. The reduced density matrix $\varrho_1$,
finally, is obtained by identifying $\bsigma_2$ and $\bsigma_2'$ and
summing, i.e. by joining the two strips between sites $(M+1)$ and $L$,
while the rest remains unconnected. Thus one arrives at the partition
function of an infinite strip with a perpendicular cut in it.

\begin{figure}[ht]
\begin{center}
\epsffile{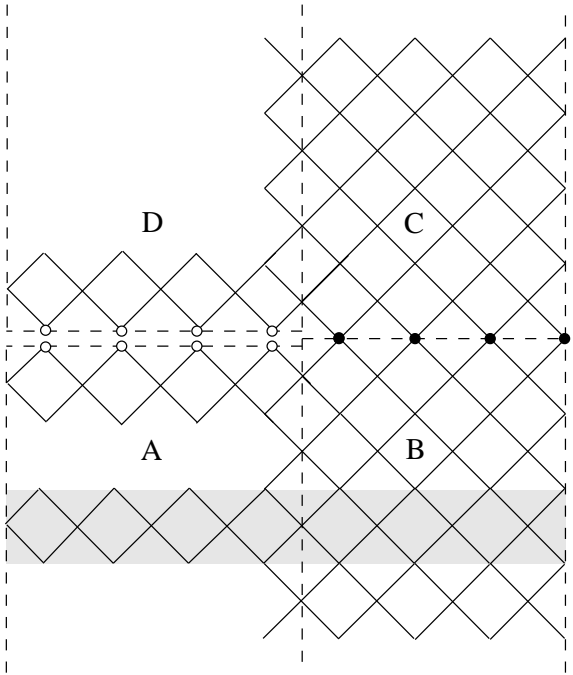}
\caption{\label{fig_ctm}The strip geometry as discussed in the text
with some portions of a square lattice. Also indicated are the corner
transfer matrices A, B, C, D and the row transfer matrix (shaded).}
\end{center}
\end{figure}

This situation, first discussed in \cite{nishino1}, is shown in Fig.
\ref{fig_ctm}. The strip in this case is formed by a diagonally
oriented square lattice. The variables $\bsigma_1$ and $\bsigma_1'$
along the cut are shown as white circles, the $\bsigma_2$ as black
ones. The dashed lines divide the system further and define four
rectangular corner transfer matrices denoted by $A$, $B$, $C$ and $D$.
They are the partition functions of the corresponding lattice pieces
and of Ramond type, i.e. without a common central spin \cite{foda}. It
follows that
\newpage
\begin{equation}
\varrho_1=ABCD.
\end{equation}
This is the basic relation between the reduced density matrix and the
CTM's \cite{nishino1,nishino2}.

\begin{figure}[ht]
\begin{center}
\epsffile{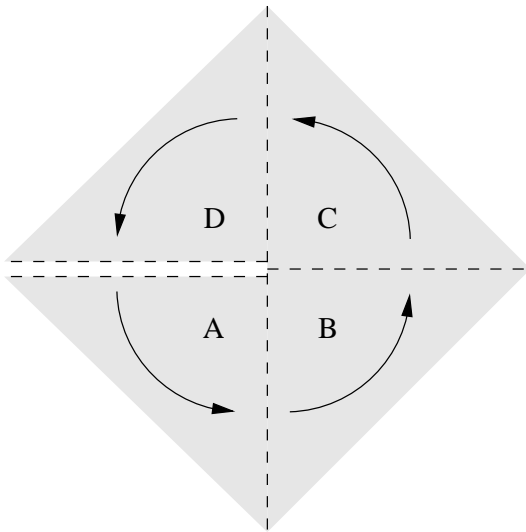}
\caption{\label{fig_ctm0} System composed of four corner transfer
matrices with standard shape. The arrows indicate the direction of the
transfer.}
\end{center}
\end{figure}

The CTM's which appear here differ somewhat from the usual ones. In
calculations for finite-size systems one normally uses the shape shown
in Fig. \ref{fig_ctm0}
\cite{baxter1,truong,truong1,peschel,davies}. The number of spins
along both edges is then the same and one is dealing with square
matrices. This is only a minor point, however, since also the products
$AB$ and $CD$ in the strip are square matrices for $M=L/2$. The main
point is that, for a large system away from criticality, the outer
boundary plays a marginal role. In the CTM spectrum it affects only
the lower end \cite{truong} and in the thermodynamic limit its effect
vanishes so that one obtains a well-defined result.

For this reason the quantity
\begin{equation}
\hat\varrho_1=\hat A \hat B \hat C \hat D
\end{equation}
where the hat denotes the infinite-lattice limit will have the same
spectrum as $\varrho_1$ corresponding to the strip, up to some deviations
at the lower end and possibly some overall shift connected with the
different normalizations of the quantities. For the infinite CTM's,
however, the spectrum is known. Basically, it follows from the
Yang-Baxter equations and the corner geometry \cite{baxter}. The relevant
results will be given in the next section.

\section{Analytical results}
\label{analytic}
For the treatment of the transverse Ising chain one considers the
square lattice of Fig. \ref{fig_ctm} with Ising spins at the lattice
points and an isotropic coupling $K$. The transfer matrix
indicated by shading in the lower part of Fig. \ref{fig_ctm}
commutes with the Hamiltonian \cite{peschel1,igloi}
\begin{equation}
\label{ham_ising}
H=-\sum_{n=1}^{L-1}\sigma_n^x-\delta \sigma_L^x -
\lambda\sum_{n=1}^{L-1}\sigma_n^z \sigma_{n+1}^z
\end{equation}
where $\delta=\cosh 2K$ and $\lambda=\sinh^2 2K$. The
enhancement of the transverse field at the right end can be neglected
for large $L$. Then one arrives at the usual homogeneous chain.

The CTM's for this Ising lattice were studied in \cite{davies1} and
are exponentials of an operator which has again transverse Ising form
but with site-dependent coefficients which increase linearly with
$n$. Due to the isotropy of the lattice all four matrices are equal
and one has
\begin{equation}
\label{rho_ctm}
\hat \varrho_1=\hat A^4=e^{-\hat H_{\rm CTM}}
\end{equation} 
where
\begin{equation}
\hat H_{\rm CTM}= c \left\{ \sum_{n\ge 1}(2n-1) \, \sigma_n^x + \lambda
\sum_{n\ge 1} 2n \, \sigma_n^z\sigma_{n+1}^z \right\}
\end{equation}
with a constant $c$ depending on $\lambda$. The
diagonalization of $\hat H_{\rm CTM}$ in terms of fermions
\cite{davies1,truong1} then leads to
\begin{equation}
\label{hctm}
\hat H_{\rm CTM}=\sum_{j=0}^\infty \varepsilon_j n_j
\end{equation}
with $n_j=c^\dagger_jc_j=0,1$ and the single-particle energies

\parbox{125mm}{\[
\qquad \qquad \varepsilon_j=\left\{
\begin{array}{rcll}
(2j+1)\varepsilon & : & \quad \lambda<1 & (T>T_{\rm c}) \\
2j\varepsilon & : & \quad \lambda>1 &(T<T_{\rm c})
\end{array} 
\right. 
\]}
\hfill
\begin{minipage}{10mm}
\begin{flushright}
\addtocounter{equation}{1} (\arabic{equation})\\
\addtocounter{equation}{1} (\arabic{equation}) 
\end{flushright}
\end{minipage}

\noindent where $\varepsilon $ is given by
\begin{equation}
\label{elliptic}
\varepsilon=\pi \frac{I(k')}{I(k)}.
\end{equation}
Here $I(k)$ denotes the complete elliptic integral of the first kind,
$k'=\sqrt{1-k^2}$ and the parameter $k$ with $0\le k \le 1$ is related
to $\lambda $ by

\parbox{125mm}{\[
\qquad \qquad k=\left\{
\begin{array}{ccl}
\lambda & : & \quad \lambda<1 \\
1/\lambda & : & \quad \lambda>1.
\end{array}
\right.
\]}
\hfill
\begin{minipage}{10mm}
\begin{flushright}
\addtocounter{equation}{1} (\arabic{equation})\\
\addtocounter{equation}{1} (\arabic{equation})
\end{flushright}
\end{minipage}

\noindent An additive constant has been omitted in (\ref{hctm}). From
this one finds the following eigenvalues $E$ for $\hat H_{\rm CTM}$.

\begin{enumerate}
\item $\lambda <1$ \label{a}

The ground state has $E=0$. Excited states with one fermion lead to
the odd levels $E=\varepsilon, 3\varepsilon, 5\varepsilon, \ldots$
With two fermions one obtains in addition all even levels {\em except}
$2\varepsilon$. The level $8\varepsilon$ is two-fold degenerate. The
three-fermion excitations start at $9\varepsilon$ and lead to further
degeneracies etc. The spacing of the $E$'s, except at $2\varepsilon$,
is therefore $\varepsilon$.

\item $\lambda >1$ \label{b}

The ground state has again $E=0$, but it is now twice degenerate due
to the long range order. This leads to a corresponding two-fold
degeneracy of all excited levels. Formally, this follows from the fact
that $\varepsilon_0=0\,$\footnote{The normalization of this state is
discussed in \cite{davies1}.}. The excited states are integer multiples
of $2\varepsilon$, i.e. the spacing of the $E$'s is $2\varepsilon$.
\end{enumerate}

This is the well-known equidistant spectrum of $\hat H_{\rm CTM}$ with
a spacing which depends on $\lambda$, i.e. on the distance from the
critical point. It follows that up to degeneracies, which will be
discussed later, $\hat \varrho_1$ has a purely exponential spectrum.

\medskip

For the Heisenberg model, the associated two-dimensional system is the
six-vertex model which is specified by three Boltzmann weights $a,b,c$
\cite{baxter}. If the vertex lattice is oriented parallel to the
strip, its row-to-row transfer matrix commutes with the Hamiltonian
\cite{sutherland} 
\begin{eqnarray}
\label{ham_hb}
H & = & \sum_n H_n \\
H_n & = & \sigma_n^x \sigma_{n+1}^x +  \sigma_n^y
\sigma_{n+1}^y + \Delta  \sigma_n^z \sigma_{n+1}^z
\end{eqnarray}
where $\Delta$ is related to the vertex weights via
$\Delta=(a^2+b^2-c^2)/(2ab)$. Strictly speaking, this holds only for
periodic boundary conditions on $T$ and $H$. As before, however, the
effect of the boundaries should not matter for a large system. The
non-critical case corresponds to $\Delta >1$, so that $H$ describes a
uniaxial antiferromagnet.

The infinite CTM's for the six-vertex model are also known
\cite{baxter,davies2,davies3}. They are again exponentials of an 
operator similar to (\ref{ham_hb}) but with site-dependent
coefficients. Thus $\hat \varrho_1$ takes the form (\ref{rho_ctm}) and
\begin{equation}
\hat H_{\rm CTM}= c\sum_{n\ge 1}n H_n
\end{equation}
where the constant $c$ depends on $\Delta$. Although the problem does
not look like a free-fermion one, the eigenvalues of $\hat H_{\rm
CTM}$ follow once more from (\ref{hctm}) and (12), with $\varepsilon$
now given by
\begin{equation}
\label{acosh}
\varepsilon={\rm arcosh} \, \Delta.
\end{equation}
Therefore the discussion for case \ref{b} above can be taken over and
one finds again an exponential spectrum for $\hat \varrho_1$.

\section{Numerical calculations}
In order to check these predictions we have carried out standard DMRG
calculations for chains with up to 100 sites, keeping between 30 and
64 states in each block. The systems were built up via the
infinite-size algorithm and $\varrho_1$ refers to one half of the
chain. In the following, the eigenvalues $\omega_n$ of $\varrho_1$,
ordered according to magnitude, are shown in semilogarithmic plots
$\log \omega_n$ vs. $n$. All cases correspond to correlation lengths
which do not exceed ten lattice spacings and thus are much smaller
than $L$.

\begin{figure}[ht]
\begin{center}
\epsfxsize=120mm
\epsfysize=100mm
\epsffile{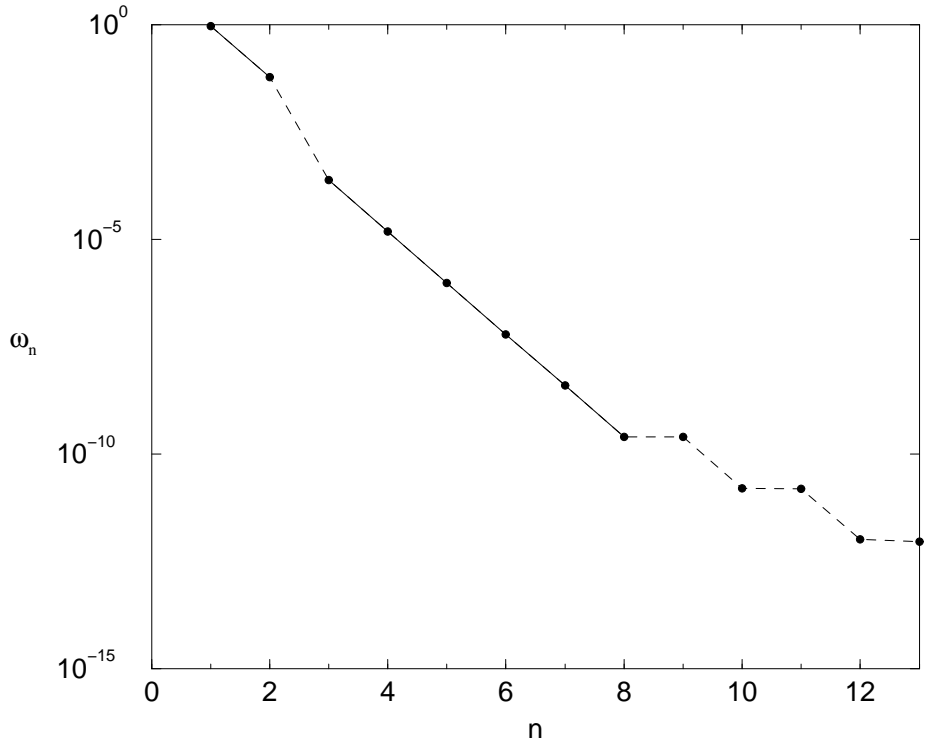}
\caption{\label{fig_is1} Density-matrix spectrum of a transverse Ising
chain with $\lambda=0.8$ and 120 sites, calculated with $30$ states.}
\end{center}
\end{figure}

We begin with results for the transverse Ising
chain. Fig. \ref{fig_is1} shows an example of the spectrum in the
disordered region $(\lambda<1)$.
One can see that the first seven levels are non-degenerate and
equidistant apart from a break between the second and the third
one. This is exactly the pattern found for the $E$'s of the infinite
lattice in Section \ref{analytic}, case \ref{a}. The straight line
through six of the levels shows clearly the linear behaviour and the
level spacing (the slope in a plot $\ln \omega_n$ vs. $n$) agrees up
to $10^{-3}$ with the theoretical value 2.7565 from
(\ref{elliptic}). At the lower end, two-fold degeneracies are
visible. Beyond that the spectrum becomes flat and no further
structure can be seen. There, however, one is already at $10^{-14}$
and near the limits of the accuracy of the calculation. For larger
values of $\lambda$, where the slope becomes smaller, the situation is
more favorable.

\begin{figure}[ht]
\begin{center}
\epsfxsize=120mm
\epsfysize=100mm
\epsffile{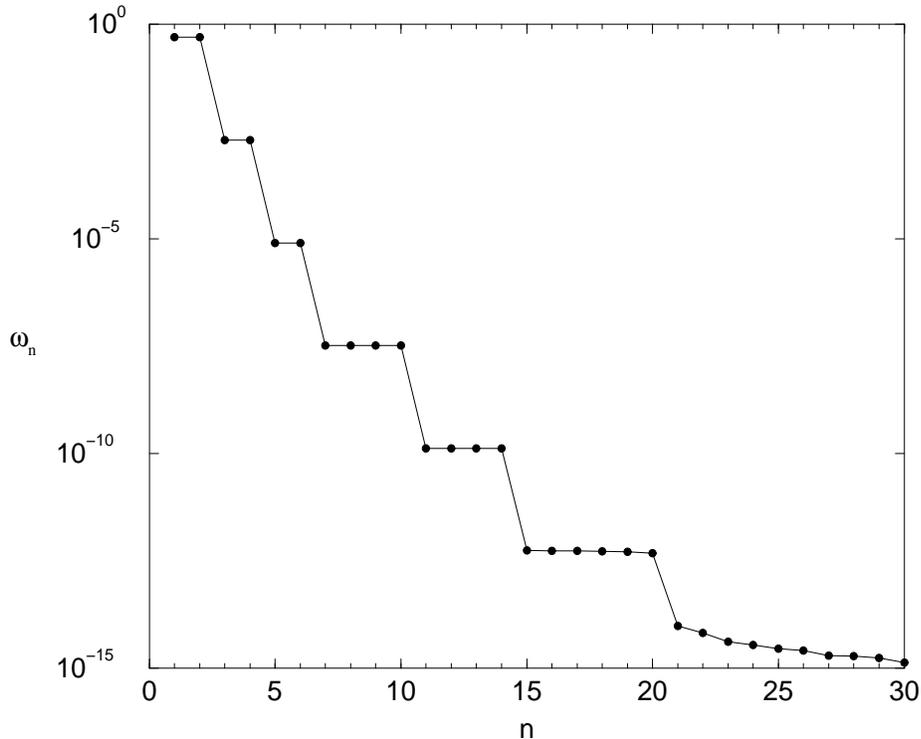}
\caption{\label{fig_is2}Density-matrix spectrum of a transverse
Ising chain with $\lambda=1.25$ and 120 sites, calculated with $30$
states.}
\end{center}
\end{figure}

The situation in the ordered phase is shown in Fig. \ref{fig_is2} for
$\lambda=1.25$.
One can see a clear difference in the structure of the spectrum: There
is no break and the multiplicities have changed. Their values
(2,2,2,4,4,6,\,\ldots) are just twice the number $P_j$ of partitions
of the integers $j=0,1,2,\ldots$ and thus correspond exactly to what
one expects according to Section \ref{analytic} \ref{b}. Only the last
multiplet is not correct. We attribute this to numerical
inaccuracies. 

The two-fold degeneracy needs some comments, though. In some
calculations for longer chains it was lost and one observed two sets
of levels (with degeneracies $P_j$) shifted with respect to each
other. This can be traced back to a mixing of the ground state and the
exponentially close first excited state of $H$ in the numerical
procedure, since at the same time the expectation value $\langle
\sigma_n^z \rangle$ became non-zero. The mechanism can be discussed
easily in the limit $\lambda \gg 1$, where
\begin{equation}
|\Phi\rangle \simeq \frac{1}{\sqrt{2}}\Big[ |\Phi_+\rangle +
|\Phi_-\rangle \Big] 
\end{equation}
with $|\Phi_+\rangle=|\uparrow\uparrow\cdots\uparrow\rangle$ and
$|\Phi_-\rangle=|\downarrow\downarrow\cdots\downarrow\rangle$ being
the two ferromagnetic states. This leads to
\begin{equation}
\varrho_1 \simeq \frac{1}{2}\Big[ |\varphi_+\rangle\langle \varphi_+ | +
|\varphi_-\rangle\langle \varphi_-| \Big] 
\end{equation} 
where the states $|\varphi_\pm\rangle$ now refer to part 1 of the
chain. Thus $\varrho_1$ has two degenerate eigenvalues $1/2$, the rest
being zero. The same holds for the density matrix constructed from the
first excited state
\begin{equation}
|\Phi'\rangle \simeq \frac{1}{\sqrt{2}}\Big[ |\Phi_+\rangle -
|\Phi_-\rangle \Big]. 
\end{equation}
However, a linear combination $a|\Phi\rangle + b|\Phi'\rangle$ with
$a^2+b^2=1$ leads to eigenvalues $(a+b)^2/2$ and $(a-b)^2/2$, i.e. to
a splitting. By targeting $|\Phi\rangle$ and $|\Phi'\rangle$ in a DMRG
calculation and forming such linear combinations one can actually
shift the two level sequences deliberately.

\begin{figure}[ht]
\begin{center}
\epsfxsize=120mm
\epsfysize=100mm
\epsffile{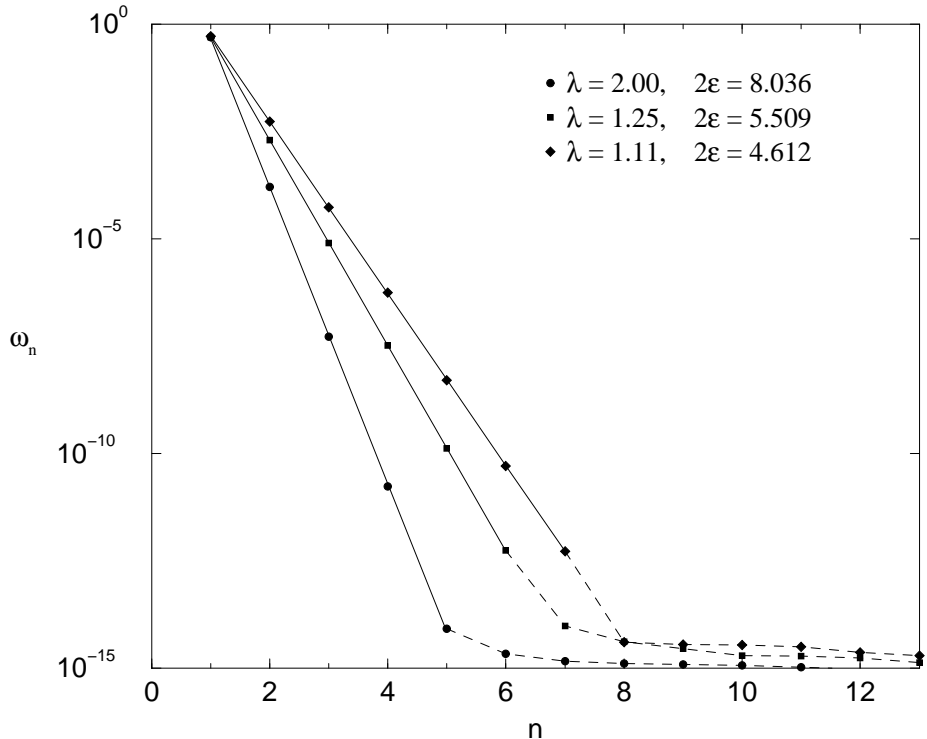}
\caption{\label{fig_is3} Density-matrix spectra of transverse Ising
chains for three values of $\lambda$ without the degeneracies,
calculated with $30$ states for $\lambda=2$ and $1.25$ and with 60
states for $\lambda=1.11$. The $\varepsilon$-values were obtained
from the slopes.}
\end{center}
\end{figure}

In Fig. \ref{fig_is3} the pure spectra without the degeneracies are
shown for several $\lambda$-values.
One sees that the slope becomes smaller near the critical point
$(\lambda=1)$, as it should, and a comparison of the numbers shows
again very good agreement with the value $2\varepsilon$ for the
spacing according to Section \ref{analytic}. Closer to the critical
point, however, a larger number of states had to be kept. Also here
the spectrum becomes flat around $10^{-14}$.

\medskip

\begin{figure}[ht]
\begin{center}
\epsfxsize=120mm
\epsfysize=100mm
\epsffile{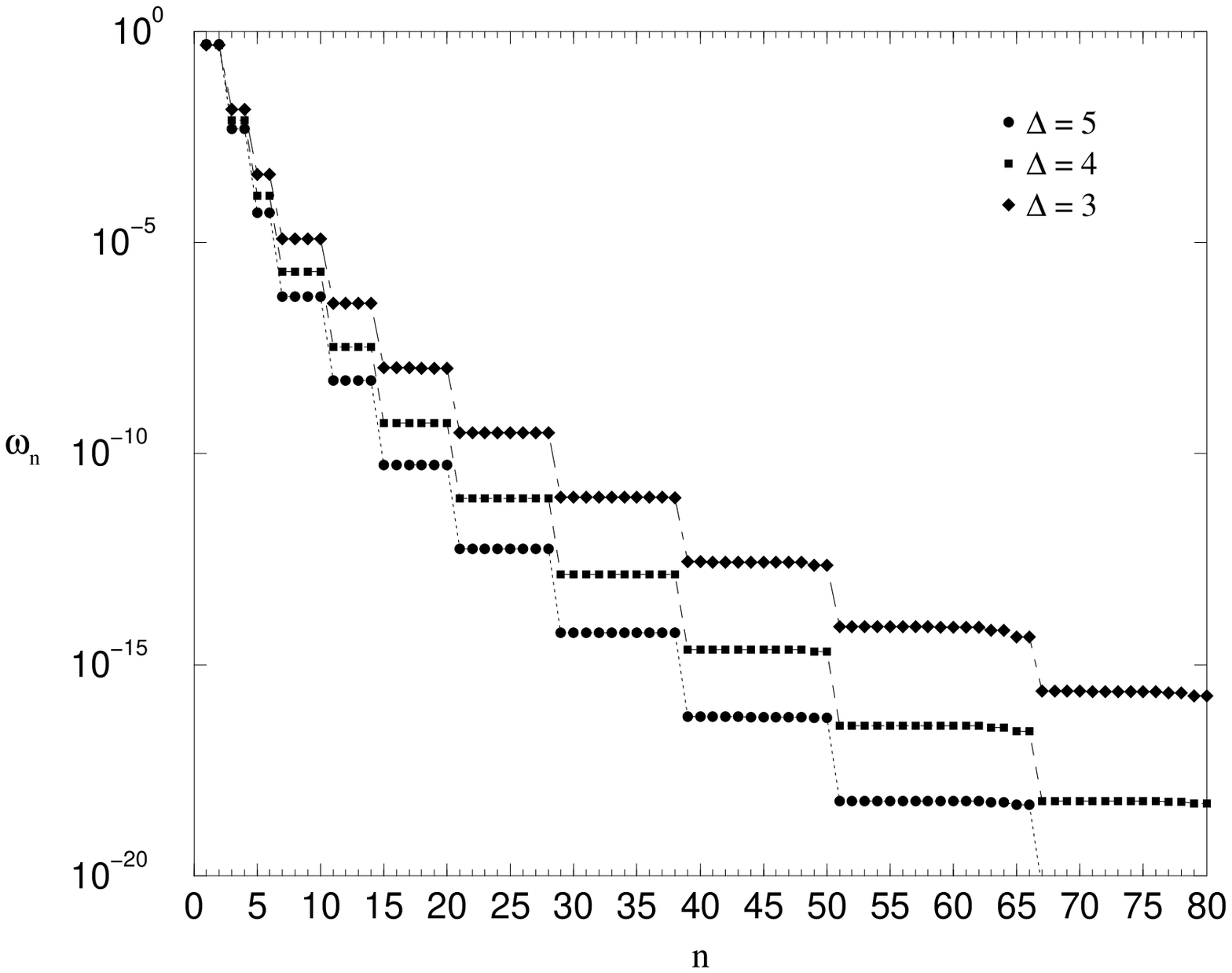}
\caption{\label{fig_hb1} Density-matrix spectra of Heisenberg chains
with 98 sites, calculated with $64$ states for three values of
$\Delta$.}
\end{center}
\end{figure}

\begin{figure}[ht]
\begin{center}
\epsfxsize=120mm
\epsfysize=100mm
\epsffile{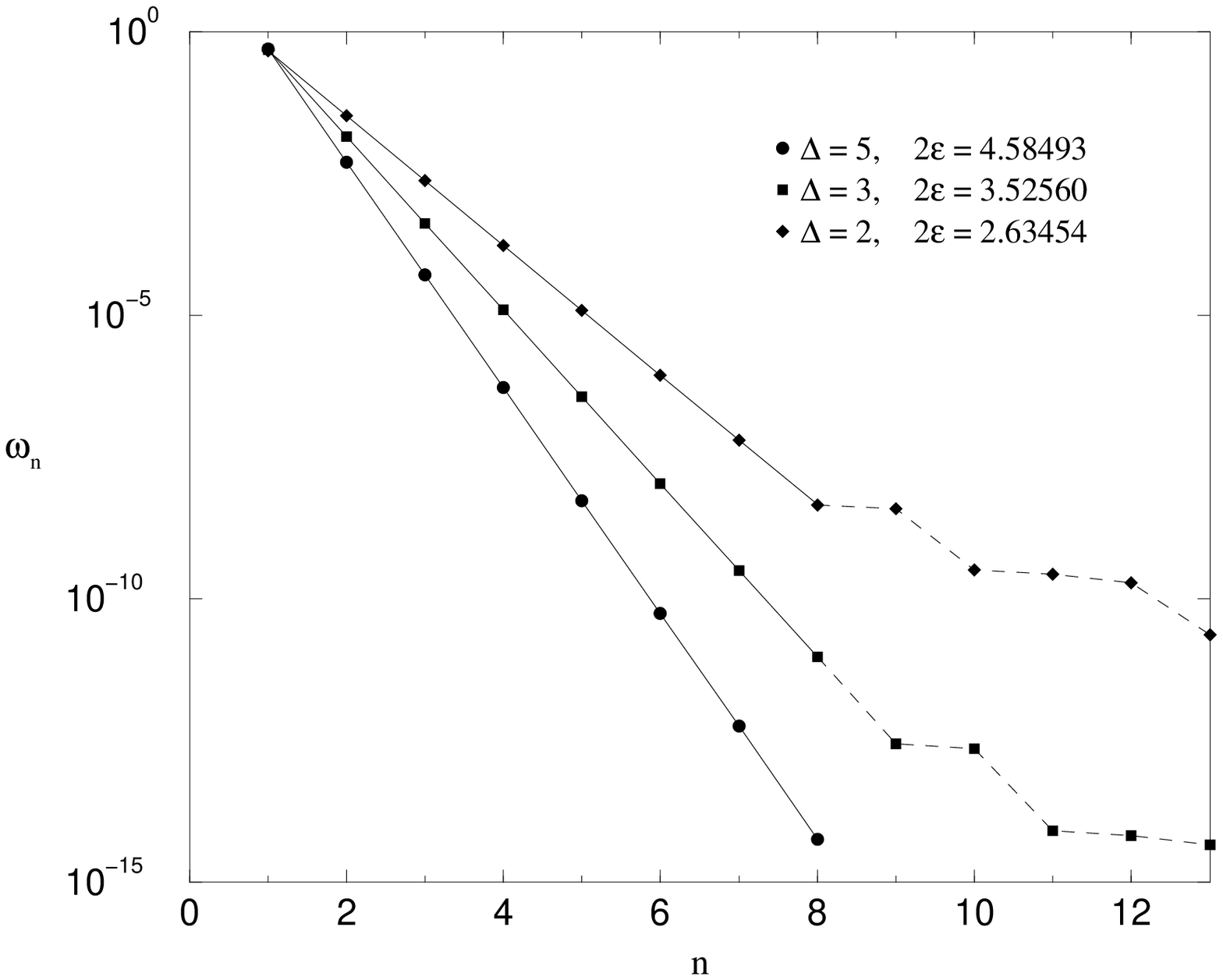}
\caption{\label{fig_hb2} Density-matrix spectra of $X\!X\!Z$ chains of
98 sites without the degeneracies, calculated with 64 states. The
$\varepsilon$-values were obtained from the slopes.}
\end{center}
\end{figure}

For the $X\!X\!Z$ chain, where only the ordered region and the
critical phase exist, the situation is even better. Here one can work
in the subspace $S^z=0$ which makes the calculation more precise with
the same number of kept states. Fig. \ref{fig_hb1} shows the complete
spectrum for three values of the anisotropy $\Delta$.
In all cases, the first 10 levels can be observed and the degeneracies
$2\cdot(1,1,1,2,2,3,4,5,6,8)$ can be read off clearly, even though
some very low levels do not fit perfectly. In comparison to
Fig. \ref{fig_is2} the accuracy is much better here so that even at
$10^{-18}$ the structures can be seen. This is a very convincing
demonstration of the theoretically predicted scheme. The perfect
two-fold degeneracy indicates that there is no admixture of the
excited state here. This is in line with the observation that
$\langle\sigma_n^z\rangle=0$ in these calculations. Finally, in
Fig. \ref{fig_hb2} pure spectra are shown together with linear
fits for the first 8 levels.
One should note that there is an arbitrariness in the definition of
degeneracy. Here the criterion was a relative difference of less than
$10^{-4}$. If one relaxes this, the linear region becomes even
larger. The $\varepsilon$-values given in the Figure agree with
(\ref{acosh}) up to 3-4 decimal places. Also here there is a
flattening of the spectrum at the lower end. This, incidentally is in
contrast to results for usual finite-size CTM's, where the spectrum
becomes steeper at the lower end \cite{truong,truong1,davies}, and
can have two origins: The different geometries of the systems
in the two calculations (Figs. \ref{fig_ctm} and \ref{fig_ctm0}) and
limitations due to the truncation in the DMRG procedure.

\section{Conclusion}
We have shown for two integrable, non-critical spin chains how the
density matrix reflects the properties of the associated corner
transfer matrices. The exponential character of its spectrum can in
these cases be understood as a consequence of the star-triangle or
Yang-Baxter equations.\footnote{It is amusing to note that such an
exponential form arises also for two coupled harmonic oscillators
\cite{han}.} The same will hold for other integrable models and one
thereby obtains a whole class of examples, where the density-matrix
spectrum can be obtained analytically. In such cases one could
estimate in advance, how many states one has to keep in a DMRG
calculation to get a certain accuracy. Since even for soluble models
the correlation functions might have to be calculated numerically,
this could be useful in practice.

One immediate question concerns, of course, non-integrable
systems. The connection to CTM's still exists, but there are no
general results for their spectra. For the three-state Potts model, it
was shown that the equidistance of the levels of $H_{\rm CTM}$ is lost
for finite temperatures \cite{wunderling}. This is consistent with
calculations of the $\varrho_1$-spectrum, which also show irregular
level spacing, although the overall behaviour is still roughly
exponential \cite{ritter}. A recent investigation of other
non-integrable models even indicates a kind of universality for the
spectra \cite{okunishi}. The fact that the basic features of the CTM's
are still similar was also used in direct renormalization calculations
for these quantities \cite{nishino2}.

The second question relates to critical systems. Actually a number of
the chains treated by DMRG belong to this category. Therefore an
understanding of the spectra for this case would be quite useful. One
does not need integrability here, instead conformal invariance may be
invoked. However, due to the infinite correlation length the shape of
the associated two-dimensional system is more important in this
case. The CTM spectra have been investigated for the usual CTM
geometry as in Fig. \ref{fig_ctm0} \cite{truong1,peschel,davies} but
one needs very large systems in order to see the linear spectrum of
$H_{\rm CTM}$ and the logarithmic size dependence $\varepsilon\sim
1/\ln L$ predicted by conformal invariance. The same seems to hold for
the strip geometry. This problem is presently still under
investigation.

\section*{Acknowledgements}
We thank D. Karevski for helpful contributions, T. Nishino and
R. Noack for various discussions and C. Ritter for making his Potts
results available. We also thank the Max-Planck-Institut f\"ur Physik
komplexer Systeme in Dresden for its hospitality.



\begin{thebibliography}{99}
\bibitem{white}
S.R. White, Phys. Rev. Lett. {\bf 69} (1992) 2863; S.R. White,
Phys. Rev. B {\bf 48} (1993) 10345

\bibitem{legeza}
\"O. Legeza, G. F\'ath, Phys. Rev. B {\bf 53} (1996) 14349

\bibitem{kaulke}
M. Kaulke, I. Peschel, Eur. Phys. J. B {\bf 5} (1998) 727

\bibitem{nishino}
T. Nishino, J. Phys. Soc. Japan {\bf 64} (1995) 3598

\bibitem{baxter}
R.J. Baxter, Exactly Solved Models in Statistical Mechanics,
Academic Press, London 1982

\bibitem{nishino1} T. Nishino, K. Okunishi, Density Matrix and
Renormalization for Classical Lattice Models, in: Srongly
Correlated Magnetic and Superconducting Systems, ed. G. Sierra,
M.A. Mart\'\i n-Delgado, Lecture Notes in Physics Vol. 478, Springer, 
Berlin, Heidelberg, 1997 p. 167 (see also cond-mat/9610107)

\bibitem{foda} O. Foda, M. Jimbo, T. Miwa, K. Miki, A. Nakayashiki,
J. Math. Phys. {\bf 35} (1994) 13

\bibitem{nishino2}
T. Nishino, K. Okunishi, J. Phys. Soc. Japan {\bf 66} (1997) 3040

\bibitem{baxter1}
R.J. Baxter, J. Stat. Phys. {\bf 15} (1976) 485; {\bf 17} (1977) 1

\bibitem{truong}
T.T. Truong, I. Peschel, Z. Phys. B {\bf 69} (1987) 385

\bibitem{truong1}
T.T. Truong, I. Peschel, Z. Phys. B {\bf 75} (1989) 119

\bibitem{peschel}
I. Peschel, T.T. Truong, Ann. Physik {\bf 48} (1991) 185

\bibitem{davies}
B. Davies, P.A. Pearce, J. Phys. A: Math. Gen. {\bf 23} (1990)
1295

\bibitem{peschel1}
I. Peschel, Phys. Lett A {\bf 110} (1985) 313

\bibitem{igloi}
F. Igl\'oi, P. Lajk\'o, J. Phys. A: Math. Gen. {\bf 29} (1996) 4803

\bibitem{davies1}
B. Davies, Physica A {\bf 154} (1988) 1

\bibitem{sutherland}
B. Sutherland, J. Math. Phys. {\bf 11} (1970) 3183

\bibitem{davies2}
B. Davies, Physica A {\bf 159} (1989) 171

\bibitem{davies3}
B. Davies, Inifinite-dimensional symmetry of corner transfer
matrices, in: Confronting the Infinite, ed. A.L. Carey et al.,
World Scientific, Singapore 1995 p. 193 

\bibitem{han}
D. Han, Y.S. Kim, M.E. Noz, cond-mat/9705029 (1997)

\bibitem{wunderling}
R. Wunderling, Diplomarbeit, Freie Universit\"at Berlin 1991

\bibitem{ritter}
C. Ritter, Private communication

\bibitem{okunishi}
K. Okunishi, Y. Hieida, Y. Akutsu, cond-mat/9810239 (1998)

\end{thebibliography}
\end{document}